\documentclass{article}
\usepackage{titling}
\predate{}
\postdate{}

\usepackage[utf8]{inputenc} % allow utf-8 input
\usepackage[T1]{fontenc}    % use 8-bit T1 fonts
\usepackage{hyperref}       % hyperlinks
\usepackage{url}            % simple URL typesetting
\usepackage{booktabs}       % professional-quality tables
\usepackage{amsfonts}       % blackboard math symbols
\usepackage{nicefrac}       % compact symbols for 1/2, etc.
\usepackage{microtype}      % microtypography
\usepackage{lipsum}
\usepackage{graphicx}
\graphicspath{ {./images/} }
\usepackage[table,xcdraw]{xcolor}
\usepackage{longtable}
\usepackage{caption}
\usepackage{subcaption}

\title{Knowledge Graph for Microdata of Statistics Netherlands}
\date{} % clear date
\author{Chang Sun \\
  Institute of Data Science\\
  Maastricht University\\
  \texttt{chang.sun@maastrichtuniversity.nl} \\
}

\begin{document}
\maketitle
% \begin{abstract}
 
% \end{abstract}

% keywords can be removed

\section{Introduction}
Statistics Netherlands (CBS)\footnote{https://www.cbs.nl/} hosted a huge amount of data not only on the statistical level but also on the individual level. They have collected and maintained data from the whole Dutch population over 100 years. With the development of data science technologies, more and more researchers request to conduct their research by using high-quality individual data from CBS (called CBS Microdata) or combining them with other data sources. CBS Microdata is linkable data at a personal, company and address level with which researchers can conduct statistical research themselves under strict conditions [1]. The requester (has to be a researcher) will have a protected working environment to store the data files,  intermediate files, and output. CBS Microdata is considered as a reliable and informative data source which covers health, socio-economic, educational, financial and other 14 categories. To a large degree, making great use of these data for research and scientific purposes can tremendously benefit the whole society. 

However, CBS Microdata has been collected and maintained in different ways by different departments in and out of CBS. The representation, quality, metadata of datasets are not sufficiently harmonized. Each dataset is briefly described in one to three sentences on website\footnote{https://www.cbs.nl/nl-nl/onze-diensten/maatwerk-en-microdata}. A more detailed description for each dataset is provided in a PDF file in Dutch on CBS website separately. Due to the lack of integration of all Microdata sets and a centralized platform to query the metadata, it is a very time-consuming and costly task for researchers to find all needed datasets or particular variables. Researchers first need to dive into all datasets description pages in the specific category. Then, researchers have to download and read (translate to English if needed) all lengthy PDF files to know the basic information about the datasets. In this way, researchers miss the relations between different datasets and are not able to easily find all needed variables across multiple datasets. Therefore, a general research question is formulated for this project: 
\textit{Can we convert the descriptions of all CBS microdata sets into one knowledge graph with high-quality and comprehensive metadata so that the researchers can easily query the metadata, explore the relations among multiple datasets, and find the needed variables?}
The above general research question can be divided into the following sub-questions:
\begin{enumerate}
\item \textit{Can we extract key information about CBS Microdata from the text (PDF files)? 
\item What are the most suitable ontologies for the CBS Microdata metadata?
\item Can we use the extracted information to make a knowledge graph on CBS Microdata metadata?
\item Can we find relations across different datasets and categories?}
\end{enumerate}

\section{Related Work}
\label{sec:headings}
Semantic web and linked data technology is not new for the statistics offices. In 2001, SDMX (Statistical Data and Metadata eXchange)  was launched to standardize and modernize the mechanisms and processes for the exchange of statistical data and metadata among international organisations and their member countries [2]. However, there are not many publications or publicly available softwares describing how to convert statistical (meta)data to a knowledge graph. EU Open Data Portal\footnote{https://data.europa.eu/} provides a SPARQL tool to query the metadata of their Linked data [3]. They created a vocabulary for the metadata using the Data Catalogue Vocabulary (DCAT)\footnote{https://www.w3.org/TR/vocab-dcat/} and Dublin Core Terms (DCT) vocabulary\footnote{https://www.dublincore.org/specifications/dublin-core/dcmi-terms/}. Sarker et al. proposed their plan and methods to implement semantic web technology for Australian Bureau of Statistics in 2017 [4]. It is a proof-of-concept paper which doesn’t provide any actual implementations. In 2018, Chaves-Fraga et al. provided a mapping translator from RMLC to R2RML and a comparative analysis over two different real statistics datasets using Data Cube Vocabulary [5]. This study focuses on converting CSV to RDF and reducing the size of the R2RML mapping documents. Existing studies only cover one part of this project.

\section{Methodology}
\label{sec:others}
\begin{figure} % picture
    \centering
    \includegraphics[width=1\textwidth]{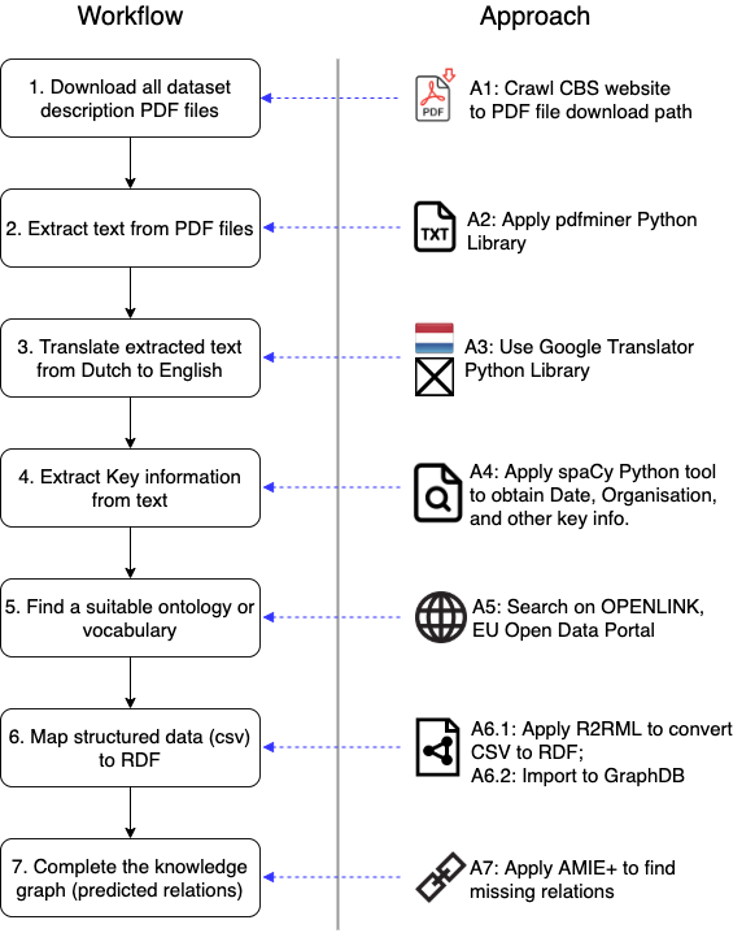}
    \caption{Workflow and approaches to building the Knowledge Graph on CBS Microdata}
    \label{fig-1}
\end{figure}

Since the descriptions of CBS Microdata are only presented by the text in PDF files, a heavy data pre-processing job is required before building up the knowledge graph. The data pre-processing tasks include extracting text from the diverse layout of PDF files, translating Dutch to English, extracting key information from the sentences, normalizing extracted information. After pre-processing, the data description text from PDF files is converted to structured data (CSV). Furthermore, I converted the CSV data to RDF and explored the knowledge graph in GraphDB. The whole workflow is presented in Figure \ref{fig-1}. This section will describe all steps in the following subsections.
\subsection{Automatically download data description files}
The data description of each dataset is presented in a PDF file which can be downloaded separately from CBS Microdata websites. To automatically download all PDF files, I wrote a Python script to crawl information from the related CBS websites and catch the downloading links of PDF files. This code is publicly available on Github repository\footnote{https://github.com/sunchang0124/KG-CBSMicrodata}.
\subsection{Extract text from PDF files}
Extracting text accurately from a PDF document is still regarded as a very challenging task. PDF was designed as an output format that gives a good viewing layout rather than a data input format. Therefore, most of the content semantics are lost when a text or word document is converted to PDF. To get a better converting result, I applied a Python package called PDFMiner\footnote{https://pypi.org/project/pdfminer/} to extract text from PDF documents. In addition to extracting pure text, it also extracts the corresponding locations, font names, font sizes, writing direction (horizontal or vertical) for each text segment. This tool has been developed and well-maintained since 2008. It is well-recognized in the text mining community because of it’s good performance.
\subsection{Translate extracted text from Dutch to English}
Many international researchers in the Netherlands are interested in CBS Microdata. However, the Microdata websites and data descriptions are both written in Dutch. One additional challenge of this project is to translate text from Dutch to English. Consider the timeframe of the project, the best option for this task is Google Translator. I applied for the Google Translator API in Python. 

\subsection{Extract key information from the text (English)}
To have a high-quality metadata, a data description which gathers all information in a text is apparently not enough. Key information such as data released date, data publisher, subject identifiers, and other metadata elements need to be extracted from the description text. Text mining techniques are required to complete this task. I applied two well-known text mining Python libraries - NLTK\footnote{https://www.nltk.org/} and spaCy\footnote{https://spacy.io/} - to recognize entities from the text. NLTK has a very good performance on part-of-speech tagging, while spaCy is outstanding on the name entity recognition. Combining two tools increased the accuracy to find key terms in the text. In the project, I focused on tagging Noun words and recognising Organization, Date, and Person. Due to the time limitation and my Dutch language level, only English text has been processed. Dutch text could be mined in the future work.
\subsection{Find suitable vocabularies}
Finding a suitable vocabulary is a key to build the knowledge graph for the metadata of CBS Microdata. As CBS is a national statistics office, I searched related ontologies and vocabularies in the statistics community. The best options from what I have observed are Data Catalogue Vocabulary (DCAT) and Dublin Core Terms (DCT) vocabulary. EU Open Data Portal also applied these two vocabularies to their metadata for the Linked Data project. 
\subsection{Convert CSV to RDF using R2RML}
After matching the vocabularies with extracted key information (metadata), I applied R2RML [6] to convert CSV (data of metadata) to RDF. I mapped the dataset, catalog, organization, variables, and keywords (of dataset) as subjects. Language tag is also used for tagging Dutch and English content. After RDF is generated successfully, all triples are imported and stored at GraphDB.
\subsection{Complete knowledge graph}
To complete the knowledge graph, I applied AMIE+ [7] to predict potential relations between entities. Additionally, I also tried a graph embedding method using Python Library Gensim. Prediction results will be discussed in the following section. 
\section{Results}
Crawling and downloading all PDF files from CBS Microdata website took less than 10 minutes including sleeping time. Sleeping time (1-5 seconds) is to avoid being detected and blocked by the website when the requests are frequently sent to the website. In total, 505 PDF documents were downloaded on 31st March 2020. 

As I discussed in the previous section, text extraction from PDF files still remains a challenge. At the end, 420 PDF documents (83.2\%) were processed successfully by PDFMiner, while 85 documents failed to be extracted to text. 420 documents were collected from 18 different categories as Table 1 shows. The two main reasons for the failure of text extraction are the unrecognizable layout  and unable to detect words and paragraphs properly. 

\begin{longtable}[c]{llll}
\hline
\rowcolor[HTML]{9FC5E8} 
NO & Category & Num of datasets & Num of variables\footnote{The number of extracted variables names from the data description files. Text extraction from PDF, language translation might cause the number of extracted variable names smaller than the actual number.} \\
\endhead
1 & Labour and social security & 114 & 766 \\
2 & Business & 34 & 198 \\
3 & Population & 49 & 300 \\
4 & Build and live & 24 & 223 \\
5 & Financial and business services & 1 & 12 \\
6 & Health and wellbeing & 62 & 444 \\
7 & Trade and catering & 3 & 36 \\
8 & Income and expenditure & 36 & 298 \\
9 & International trade & 1 & 1 \\
10 & Industry and energy & 8 & 60 \\
11 & Agriculture & 2 & 13 \\
12 & Macroeconomy & 0 & 0 \\
13 & Nature and environment & 2 & 14 \\
14 & Education & 35 & 391 \\
15 & Government and politics & 4 & 3 \\
16 & Prices & 5 & 52 \\
17 & Security and justice & 28 & 237 \\
18 & Traffic and transport & 6 & 44 \\
19 & Leisure and culture & 6 & 23 \\
\multicolumn{2}{l}{In total} & 420 & 3115 \\
\hline
\caption{Checklist for reporting PPDDM studies}
\label{tab:my-table}
\end{longtable}

Google Translator API took around 5 minutes to translate all text from Dutch to English with acceptable results. I evaluated the performance by checking a random sample of translated text. Most of the sentences and terms were translated correctly except the ones that were broken from the text extraction. Unfortunately, I am not able to check the translation accuracy on a large scale by myself. This could be done in the future work. 
After translation, 125 key terms such as Dates, Organizations, Persons, Topics were extracted from the English text using NLTK and spaCy. Some entities were mislabelled in the results which required manual correction. For instance, the full name and abbreviation of some organizations both appear in the data description which cannot be recognized by spaCy. Another example is the label names of some variables were mislabelled as Persons or Organizations.  

\begin{figure}[h]
     \centering
     \begin{subfigure}[b]{1\textwidth}
         \centering
         \includegraphics[width=1\textwidth]{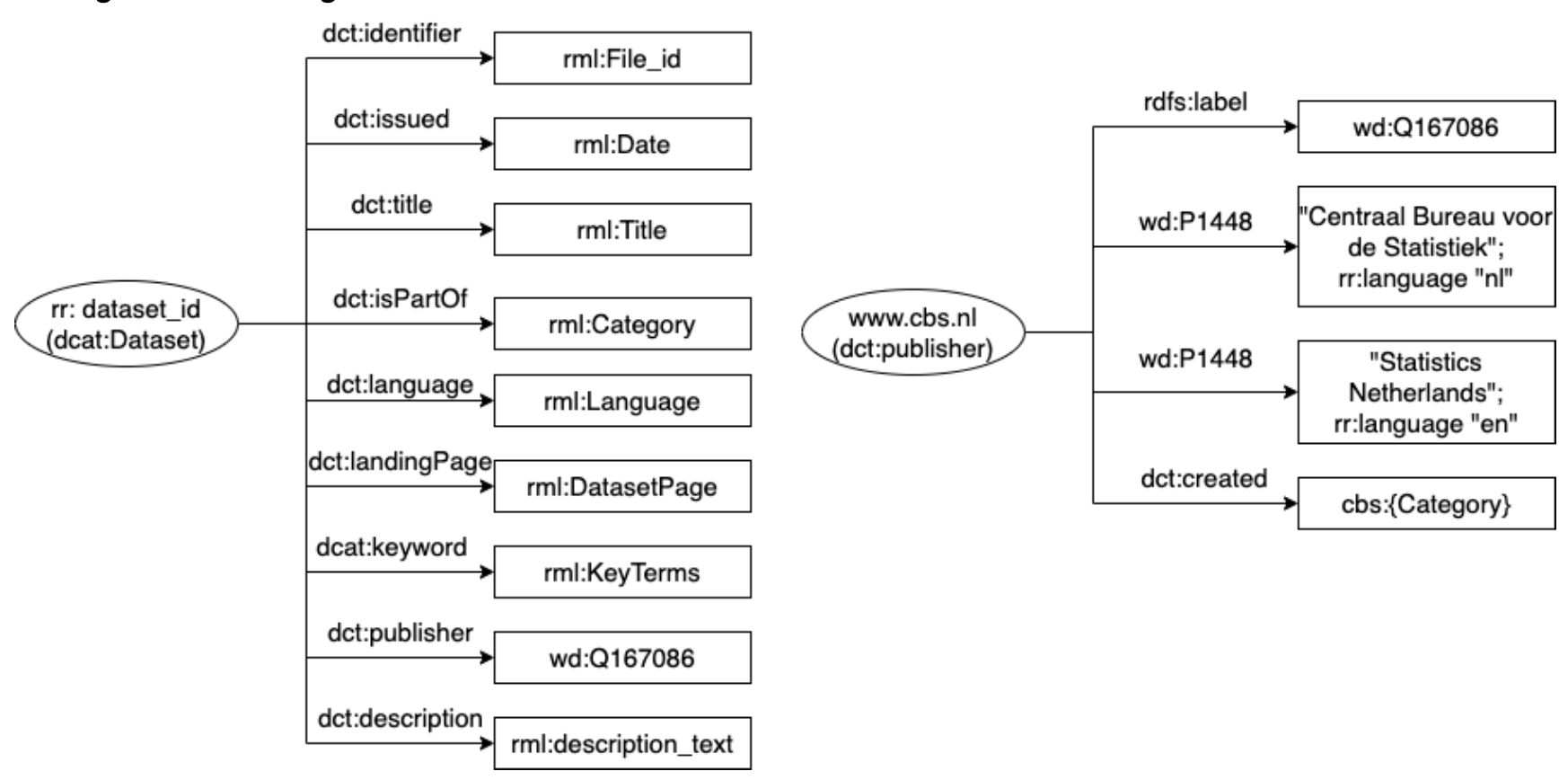}
          \caption{R2RML triple mapping diagrams for “dataset” and “publisher”}
     \end{subfigure}
     \vfill
     \begin{subfigure}[b]{1\textwidth}
         \centering
         \includegraphics[width=1\textwidth]{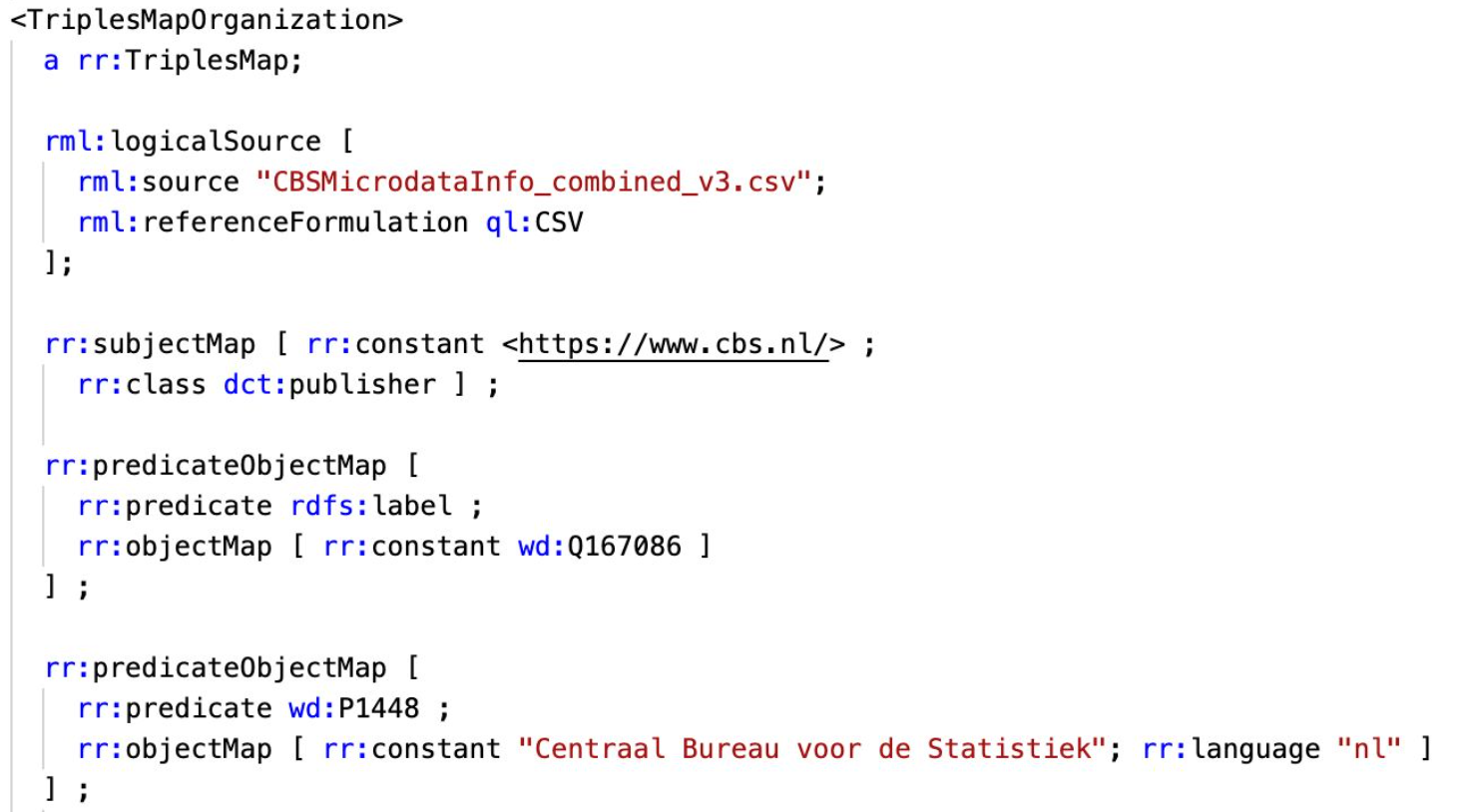}
         \caption{R2RML triple mapping for “publisher” entity}
     \end{subfigure}
\caption{R2RML mapping examples}
\label{fig-2}
\end{figure}

\begin{figure}[h] % picture
    \centering
    \includegraphics[width=1\textwidth]{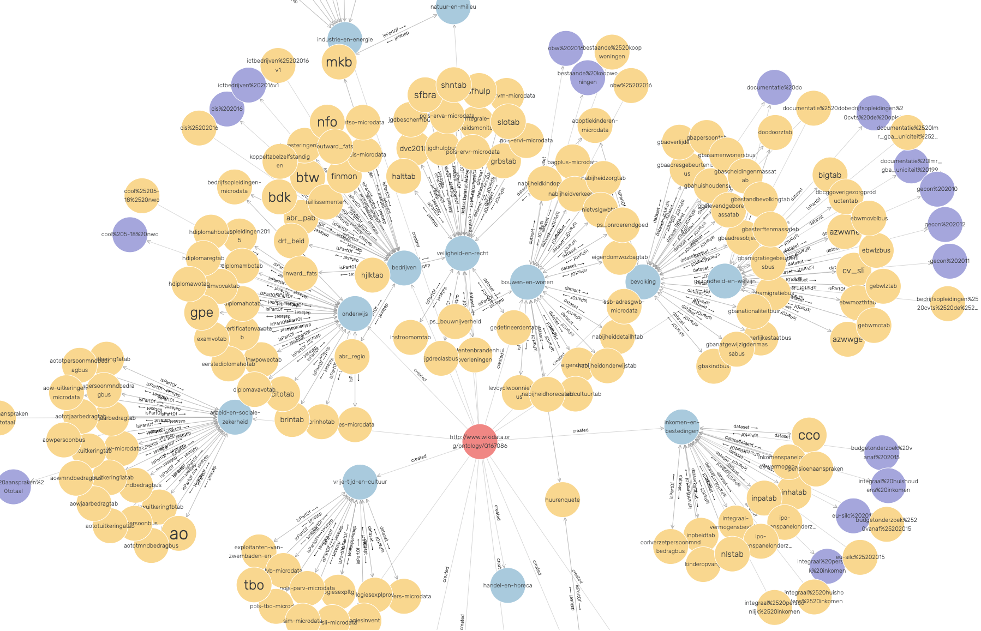}
    \caption{Visualise graph based on a part of created triples}
    \label{fig-3}
\end{figure}

To convert to RDF data, 5 triple maps were created for the dataset, catalog, organization, variables and keywords using DCAT and DCT vocabularies in the R2RML mapping file. Due to the page limitation, only the triple maps of the dataset and publisher are presented in this report (Figure \ref{fig-2}). The whole mapping file can be found in the Github repository\footnote{https://github.com/sunchang0124/KG-CBSMicrodata}. As Figure \ref{fig-2} shows, some elements of metadata such as dct:issued (Date of formal issuance (e.g., publication) of the item), dct:title, dct:description, dct:identifier, dct:language, dct:isPartOf, dct:langingPath, dcat:keyword, dct:publisher, dct:creator can be fulfilled by the extracted key information. At the end, 20242 triples were created by R2RML within 8 seconds and imported to GraphDB (Figure \ref{fig-3}). In GraphDB, some relations can be easily found by querying the knowledge graph, but they are not discoverable in the 420 lengthy PDF documents. For example, I found 251 out of 420 datasets share some variables. 43 datasets belong to two categories at the same time, while 4 datasets belong to three categories.

% \begin{figure} % picture
%     \includegraphics[width=1\textwidth]{Fig_3.png}
%     \caption{R2RML triple mapping diagrams for “dataset” and “publisher”}
%     \label{}
% \end{figure}
% \begin{figure} % picture
%     \centering
%     \includegraphics[width=1\textwidth]{Fig_4.png}
%     \caption{R2RML triple mapping for “publisher” entity}
% \end{figure}

In the last step, I applied AMIE+ to predict potential relations between entities. As this knowledge graph is not very complicated, only 9 rules were found by AMIE+. For instance, ?b  <http://purl.org/dc/terms/hasPart>  ?a   => ?a  <http://purl.org/dc/terms/isPartOf>  ?b. I insert 4 out of 9 rules to the existing graph based on their confidence score. In addition, I also tried a graph embedding method using Gensim. This method can find two datasets which are similar to each other but in two different categories because they share the same keywords or variable names. However, since information might be lost in the text extraction and translation steps,  it’s not very convincing to add new relations based on the similarity in this case. 

\section{Conclusion}
This project converts the descriptions of all CBS microdata sets into one knowledge graph with comprehensive metadata in Dutch and English.  Researchers can easily query the metadata, explore the relations among multiple datasets, and find the needed variables. For example, if a researcher searches a dataset  about “Age at Death” in the Health and Well-being category, all information related to this dataset will appear including keywords and variable names. “Age at Death” dataset has a keyword - “Death”. This keyword will lead to other datasets such as “Date of Death”. “Cause of Death”, “Production statistics Health and welfare” from Population, Business categories, and Health and well-being categories. This will tremendously save time and costs for the data requester but also data maintainers. 
However, there are some limitations in this short-term project. Firstly, only 83.2\% PDF documents were extracted to text due to several reasons such as different versions of PDF files, and unrecognizable layout. Second, accuracy of language translation and entity recognition need to be evaluated on a larger scale and optimized. For example, several dates can be extracted from the data description of one dataset. The dates might be the data collecting time, publishing time, or modifying time. More information needs to be accurately extracted from the data description documents and map to the metadata vocabulary.

\bibliographystyle{unsrt}  


\end{document}